\documentclass[9pt,twocolumn,twoside]{opticajnl}
\journal{opticajournal} 

\setboolean{shortarticle}{true}

\usepackage{graphicx}
\usepackage{stfloats}
\usepackage{float}

\title{Observation of single-photon azimuthal backflow with weak measurement}

\author[1]{Zhen-Fei Zhang}
\author[1]{Peng-Fei Huang}
\author[1]{Shan-Chuan Dong}
\author[1]{Yan-Xin Rong}
\author[2,3,4]{Jin-Shi Xu}
\author[1]{Yong-Jian Gu}
\author[1,*]{Ya Xiao}
\affil[1]{College of Physics and Optoelectronic Engineering, Ocean University of China, Qingdao 266100, People's Republic of China}
\affil[2]{CAS Key Laboratory of Quantum Information, University of Science and Technology of China, Hefei 230026, China}
\affil[3]{CAS Center For Excellence in Quantum Information and Quantum Physics, University of Science and Technology of China, Hefei 230026, China}
\affil[4]{Hefei National Laboratory, University of Science and Technology of China, Hefei 230088, China}

\begin{abstract}

Quantum backflow, a counterintuitive interference phenomenon where particles with positive momentum can propagate backward, is important in applications involving light-matter interactions. To date, experimental demonstrations of backflow have been restricted to classical optical systems using techniques like slit scanning or Shack-Hartmann wavefront sensing, which suffer from low spatial resolution due to the inherent limitations in slit width and lenslet array density. Here, we report an observation of azimuthal backflow both theoretically and experimentally by employing the weak measurement technique, which enables the precise extraction of photon momentum at each pixel. Our results show that a heralded single photon, prepared in specific superposition states with solely negative orbital angular momentum (OAM), can exhibit positive OAM. The effects of mode ratio, propagation distance and OAM index on the azimuthal backflow are systematically investigated. This work provides new techniques for observing and manipulating backflow in quantum systems. 

\end{abstract}

\setboolean{displaycopyright}{false} 

\begin{document}

\maketitle

\section{Introduction}
Quantum backflow (QB) is a counterintuitive phenomenon where a free quantum particle with positive momentum can appear to propagate backward. This phenomenon occurs in any system that supports coherent wave interference, including quantum wave functions and classical waves. Although first discovered  by Allcock in 1969 \cite{Allcock1969-1,Allcock1969-2,Allcock1969-3}, QB was not thoroughly investigated until 1994, when Bracken and Melloy proposed a novel dimensionless constant to quantify the probability of observing backflow \cite{Bracken1994}. Since then, numerous endeavors have been made to enhance the probability of observing QB through both  numerical estimations \cite{eveson2005,penz2006} and rigorous analytical solutions \cite{yearsley2012,halliwell2013,berry2010,strange2024,miller2021a}. It has been demonstrated that for particles moving along a line, the QB phenomenon occurs with a maximum probability of about 4$\%$ \cite{Bracken1994}. However, this probability can be increased either through many-wave superpositions \cite{berry2010} or extended to two-dimensional scenarios \cite{goussev2021, strange2012, paccoia2020, barbier2023a}. For instance, if charged particles move in a uniform magnetic field in an infinite two-dimensional plane \cite{strange2012, paccoia2020} or on a finite disk with a magnetic flux line passing through its center \cite{barbier2023a}, the QB probability can be unbounded.  Recently,  Trillo \textit{et  al.} have shown that the QB probability can also be seen as a measure of maximum quantum advantage \cite{trillo2023}.

The QB phenomenon has also been investigated in many-particle systems \cite{barbier2023a} and relativistic systems \cite{su2018, ashfaque2019,dibari2023}, as well as in scenarios involving constant force \cite{melloy1998}, spin-orbit coupling \cite{mardonov2014}, thermal noise \cite{albarelli2016},  particle decay \cite{vandijk2019, goussev2019}, dissipations \cite{mousavi2020}, and scattering  \cite{bostelmann2017}. Connections between QB and various phenomena, such as the arrival-time problem \cite{muga1999, muga2000, halliwell2019}, Bose-Einstein condensate \cite{palmero2013}, superoscillation \cite{berry2010}, and quantum reentry \cite{goussev2020}, have also been explored. QB is a manifestation of rapid phase changes and has potential applications in trapping micro-particles \cite{gahagan1999,li2020,tian2021,geint2022}, improving imaging resolution
  \cite{zheludev2021b}, and enhancing chiral light-matter interactions \cite{tang2011}.  

Despite significant theoretical advancements in QB, experimental demonstrations remain limited. To date, only three instances of backflow have been experimentally observed: two involving one-dimensional transverse momentum backflow \cite{eliezer2020,daniel2022} and one involving two-dimensional transverse momentum backflow, also known as azimuthal backflow (AB) \cite{ghosh2023}. The transverse momentum was measured using the slit scanning technique \cite{eliezer2020} or the Shack-Hartmann wavefront sensor (SHWFS) technique \cite{daniel2022,ghosh2023}. However, a wider slit reduces the probability of observing  backflow, while a narrower slit causes diffraction, resulting in low spatial resolution. SHWFS retrieves the transverse momentum by measuring different wavefront derivatives, whose spatial resolution is limited by the density of the Fourier transform lenslet array. Given the fragile nature of the QB phenomenon and the limitations of current detection techniques, all existing experimental demonstrations have been conducted in classical optical systems. There is an urgent need for novel detection techniques with enhanced spatial resolution to experimentally observe QB in quantum systems.

Here, we investigate the AB phenomenon within a single-photon system. A triggered single photon was generated via spontaneous parametric down-conversion and subsequently prepared in specific superposition states of two Laguerre-Gaussian(LG) modes with solely negative OAM indices. By performing a weak measurement of the photon's momentum followed by a strong measurement of its position, the azimuthal momentum of the photon at each pixel can be extracted.  Our results reveal that positive azimuthal momentum appears in regions of destructive interference where the interference visibility exceeds the threshold, indicating the presence of AB. The distribution of AB can be flexibly manipulated by changing the mode ratio, propagation distance, and OAM index.
Our work represents a step towards the observation and manipulation of backflow in quantum systems, where the flexible manipulation of the backflow region has potential applications in the optical trapping of microparticles and array optical tweezers.

\section{Theory}

The wave-function of a single photon in the paraxial LG mode with OAM index $l$ and radial index  $p$  
can be written as \cite{allen1992}
\begin{equation}
	\begin{split}
		&u_{l,p}(r,\phi ,z) = \sqrt {\frac{2p!}{\pi (p + \left| l \right|)!}} \frac{1}{\omega (z)}\left(\frac{\sqrt 2 r}{\omega (z)}\right)^{\left| l \right|}
		 L_p^{\left| l \right|}\left(\frac{2r^2}{\omega (z)^2}\right)  \\
   &\times \exp (- \frac{r^2}{\omega (z)^2}) \exp {\left( \frac{i k r^2 z}{2 (z^2 + {z_R}^2)} +i kz + il\phi  -i \psi (z)\right)},
	\end{split}
	\label{SLG}
\end{equation}
where $\left({r,\phi}\right)$ are the polar coordinates in the transverse plane, 
$\omega(z) = \omega_{0} \sqrt {1 + z^2/z_R^2} $ is the beam waist at distance $z$, $k$ is the wave number, 
${z_R} = k\omega_0^2/2$ is the Rayleigh range, 
${L_p}^{\left| l \right|}(\cdot )$ is the associated Laguerre polynomial and $\psi (z) = (2p + \left| l \right|)\arctan ( z/z_R)$ is the Gouy phase shift.

For simplicity, we are interested in the superposition of two LG modes with different OAM indices $l_1$ and $l_2$, and identical radial indices $p_1=p_2=0$, which can be expressed as
\begin{equation}
\Phi_{s} (r,\phi,z) = {u_{{l_1},0}}(r,\phi,z) + b{u_{{l_2},0}}(r,\phi,z),
\label{TLG}
\end{equation}
where $l_{1} $ and $l_{2}$ are both negative or positive,  $b\in(0,1]$ is the ratio of these two LG modes. To demonstrate the existence of AB, it is necessary to calculate the azimuthal component of the photon's wavevector or momentum. For a single photon in the superposition state described in Eq. (\ref{TLG}), its azimuthal wavevector can be expressed as  \cite{ghosh2023}

\begin{equation}
	\begin{split}
		&k_{\phi,s} = \frac{1}{r}\frac{\partial }{\partial \phi }\Phi_{s} (r,\phi ,z),\\
  &=\frac{1}{2r}[ l_{1} + l_{2} + \frac{\bigl( l_{1} - l_{2} \bigr) \bigl( 1-f(b,r,l_{1},l_{2})^{2} \bigr) }{1+f(b,r,l_{1},l_{2})^{2} +2f(b,r,l_{1},l_{2}) h(l_{1} ,l_{2}, \phi, z) } ],
	\end{split}
	\label{kphi}
\end{equation}
where $h(l_{1} ,l_{2}, \phi, z)=\cos[(l_{1} -l_{2} )\phi +(\left | l_2 \right | -\left | l_1 \right | ) \arctan (z/z_R)]$ and $f(b,r,l_{1},l_{2} )=b\sqrt{\left | l_{1} \right |! / \left | l_{2} \right |!} (\sqrt{2}r /\omega(z) )^{\left | l_{2}  \right | -\left | l_{1}  \right | } $ . It is clear that the azimuthal wavevectors $k_{\phi,l_1}$ and  $k_{\phi,l_2}$ of the constituents LG mode have a constant direction at any radius: clockwise (counterclockwise) for negative (positive) signs of $l_1$ and $l_2$. However, as seen from Eq. (\ref{kphi}), $k_{\phi,s}$ has the potential to point in the counterclockwise (clockwise) direction at some given radius, depending on $\phi$ and $b$, thus indicating the existence of AB. The relation among the AB distribution, phase singularity, OAM index, and propagation distance, as well as the threshold of interference visibility for observing AB phenomenon are presented in Sections 1 and 2 of Supplement 1.

\section{Experiment}

\begin{figure}[!hbtp]
	\centering
	\includegraphics[width=0.7\linewidth]{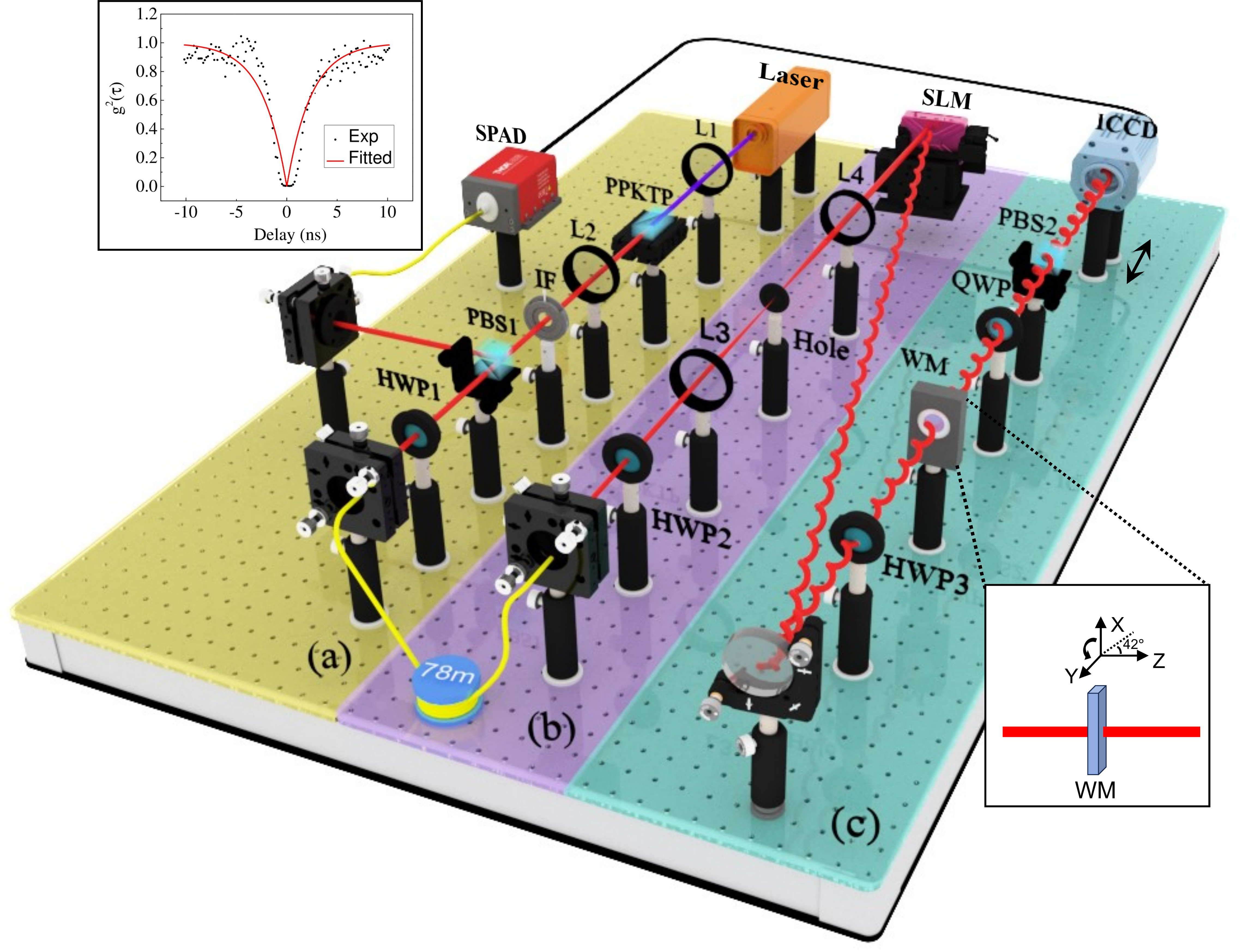}
	\caption{Experimental setup: (a) Heralded single-photon source preparation setup; (b) Superposition state of two LG modes preparation setup; (c) Wavevector measurement setup. The  top-left inset shows the second-order photon correlation function of the emitted photons. $g^2(0) = 0.006$ (with the fitting to 0.004) clearly confirms that it is a good-quality single-photon source. The bottom-right inset shows the optical axis orientation of the weak measurement calcite. The $x(y)$-component of wavevectors can be obtained by setting the optical axis of the weak measurement (WM) crystal in the $x(y)-z$ plane, respectively. The propagation distance can be adjusted by moving the intensified charge-coupled device along the $z$-direction. }
	\label{setup}
\end{figure}

The detailed experimental setup is shown in Fig. \ref{setup}. A 405-nm laser is focused on a 15 mm long type-II periodically poled KTiPO4 (PPKTP) crystal through a lens (L1) to generate idler and signal photons through the spontaneous parametric down-conversion (SPDC). These photons are then collimated with another lens (L2) and filtered by an interference filter (IF) centered at 810 nm with a bandwidth of 3 nm. After separating them with a polarization beam splitter (PBS), the idler photon is directly detected by a single photon avalanche detector (SPAD). A click on the SPAD triggers the existence of a signal photon. To verify the character of the single-photon source, the signal photon is first sent to a Hanbury Brown-Twiss interferometer, as detailed in Section 3 of Supplement 1. The second-order photon correlation function at zero delay is $g^2(0) = 0.006$, with a fitted value of $0.004$,  which demonstrates that it is a good-quality single-photon source (the top-left inset in Fig. \ref{setup}).

The detected electronic signal is used as a trigger for the intensified charge-coupled device (ICCD) camera that images the signal photon. To ensure timely detection, the signal photon is delayed by a 78-m-long single-mode fiber. Subsequently, the signal photon is expanded and collimated by lenses with focal distances $L3 = $ 75 mm and $L4 = $ 100 mm. Adopting the technique discussed in \cite{ bolduc2013}, any state described in Eq. (\ref{TLG}) can be prepared by a phase-only spatial light modulator (SLM). Since the SLM is polarization-sensitive, a pair of half-wave plates (HWP1 and HWP2) is employed to modify the polarization and optimize the diffraction efficiency.

\begin{figure*}[!ht]
	\centering
	\includegraphics[width=0.88\linewidth]{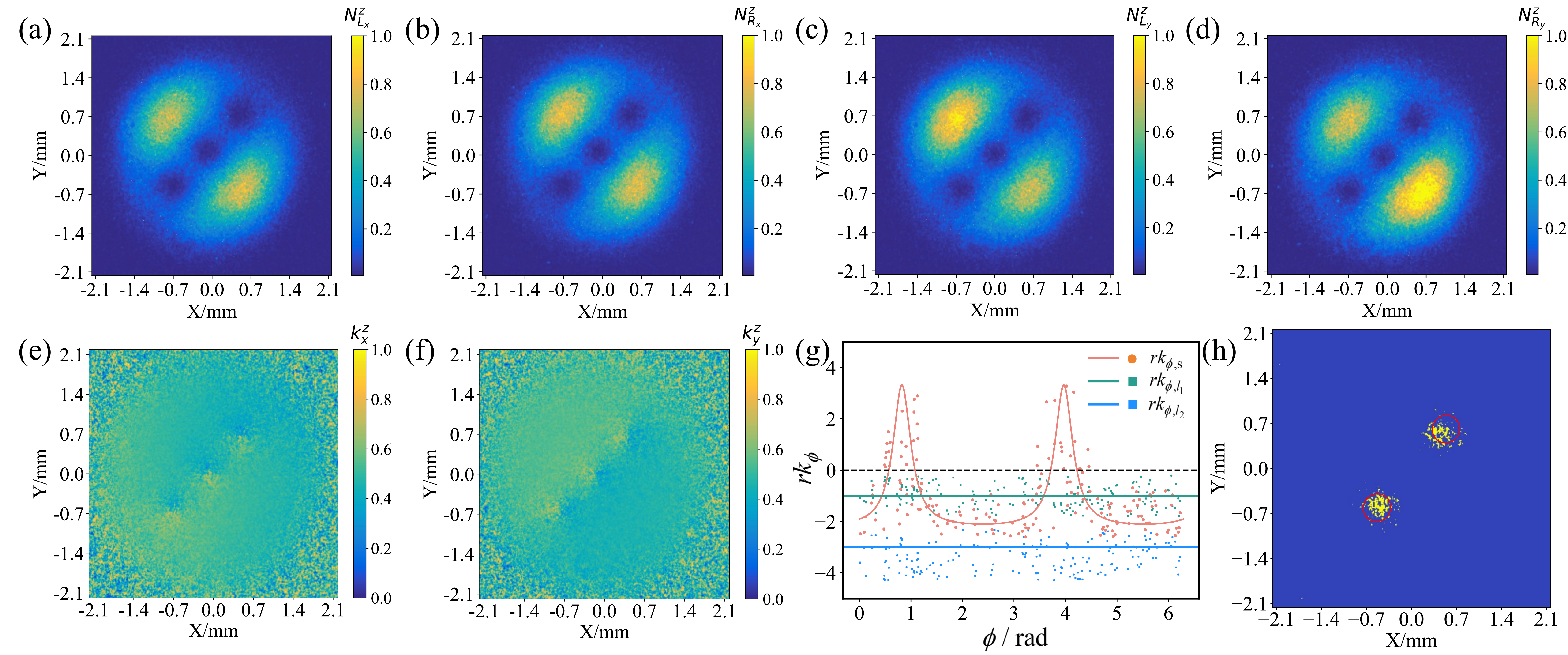}
	\caption{Experimental observation of AB with the following parameters: $l_{1} = -1$, $l_{2} = -3$, $b = 1$, and $z = 1755$ mm. (a/b) and (c/d) represent the photon counts corresponding to left/right-hand circular polarization when the optical axis of the weak measurement crystal is set in the $x-z$ plane and $y-z$ plane, respectively. (e) and (f) represent the $x$-component and $y$-component of the wave-vectors. (g) The azimuthal wave vectors $rk_{\phi}$ for each constituent (green, blue constant lines) and their superposition (orange curve) at a fixed radius $r=0.94$ mm as functions of the azimuth angle $\phi$. Experimental results and theoretical predictions are represented as circles and curves, respectively. (h) The distribution of AB. The yellow shaded regions represent the experimental AB regions, while the regions enclosed by red curves represent the theoretical AB regions. }
	\label{procedure}
\end{figure*}

The signal photon is then prepared in the diagonal polarization state by HWP3 and directed to a 0.58-mm-thick piece of calcite for weak interaction.  To measure the weak value of the wavevector $\langle k_{x(y)}^{z}\rangle_{w}$ along the $x$ ($y$) axis, the optical axis of the calcite is set in the {\it x-z} ({\it y-z}) plane, which is oriented at  $42^{\circ}$ to the $z$ axis.  This induces a small rotation in the polarization state, dependent on the wave vector  $ (\left | H  \right \rangle  +\left | V \right \rangle)/\sqrt{2} \rightarrow (\left | H  \right \rangle  +e^{i \zeta k_{x(y)}/|k|} \left | V \right \rangle)/\sqrt{2}$. The dimensionless coupling strength, $\zeta = 413$, is determined using the method detailed in our previous work \cite{xiao2017}. The rotation is then measured by a quarter wave plate (QWP) and the PBS2, with the optical axis of the QWP  set to $+45^{\circ}$/$-45^{\circ}$, which projects the final polarization state into the right/left-hand circular basis. For a given propagation distance $z$, the photon's position is recorded by a ICCD camera, which consists of 1600 $\times$ 1088 pixels and the length of each square pixel is 9 $\mu$m. The weak value of the corresponding wave vector $\langle k_{x(y)}^{z}\rangle_{w}$ is obtained from many runs by 

\begin{equation}
\langle k_{x(y)}^{z}\rangle_{w}=\dfrac{|k|}{\zeta} \arcsin\left(\dfrac{N^{z}_{R_{x(y)}}-N^{z}_{L_{x(y)}}}{N^{z}_{R_{x(y)}}+N^{z}_{L_{x(y)}}}\right).
	\label{transverse}
\end{equation}
$N^{z}_{R_{x(y)}}$ and $N^{z}_{L_{x(y)}}$ are the photon counts corresponding to the right-hand and left-hand circular polarization, respectively \cite{kocsis2011,xiao2017, xiao2019,yang2020}. The azimuthal wave-vector $k_{\phi}^{z}$ at each pixel ($x_{z}, y_{z}$) can be expressed as
\begin{equation}
	 k_{\phi }^{z}   =\frac{x_{z}\times \left \langle k_{y}^{z}  \right \rangle _{w} - y_{z}\times\left \langle k_{x }^{z } \right \rangle _{w}}{\sqrt{x_{z}^2+y_{z}^2}}.  
	\label{azimuthal}
\end{equation}

Firstly, the single photon is prepared in the superposition state of two LG modes with $l_{1} = -1 $, $l_{2} = -3$ and $b = 1$. We take this as an example to present the concrete procedures for quantifying the AB using the experimentally observed probability density distribution. The results for a propagation distance of $z = 1755$  mm are illustrated in Fig. \ref{procedure}. The photon counts corresponding to right-hand circular polarization ($N^{z}_{R_{x}}$, $N^{z}_{R_{y}}$) and left-hand circular polarization ($N^{z}_{L_{x}}$, $N^{z}_{L_{y}}$) are depicted in Fig. \ref{procedure}(a-d), respectively. According to Eq. (\ref{transverse}), we can iteratively calculate the normalized $x$-component and $y$-component of wavevectors, as shown in Fig. \ref{procedure}(e) and (f). Furthermore, the normalized azimuthal components $k_{\phi,s}/k$  of wave-vectors can be obtained using Eq. (\ref{azimuthal}). In Fig. \ref{procedure}(g), we present a quantitative analysis of the relation  between the azimuthal components of the wavevectors ${ rk_{\phi,l_1}, rk_{\phi,l_2}, rk_{\phi,s} }$ and the azimuthal angle $\phi$ at a fixed radius of $r=0.94$ mm. Here, $k_{\phi,l_1}$, $k_{\phi,l_2}$, and $k_{\phi,s}$ represent the azimuthal wavevectors for LG mode $l_1$, LG mode $l_2$, and their superposition, respectively. Clearly, when $l_{1}<0$ and $l_{2}<0$, then $rk_{\phi,l_1}<0$ and $rk_{\phi,l_2}<0$, which implies $rk_{\phi,s}>0$, indicating the existence of AB. The distribution of AB regions is presented in Fig. \ref{procedure}(h). Yellow (Blue) regions correspond to experimental values of $rk_{\phi,s}$ being (not) larger than zero, indicating (non-)appearance of AB. The regions enclosed by the red curves correspond to the theoretical AB regions. The discrepancy between experimental results and theoretical predictions primarily stems from the limited number of photons in the destructive interference region, rendering them particularly vulnerable to the influences of environmental noise.

\begin{figure}[tb!p]
	\centering
	\includegraphics[width=0.9\linewidth]{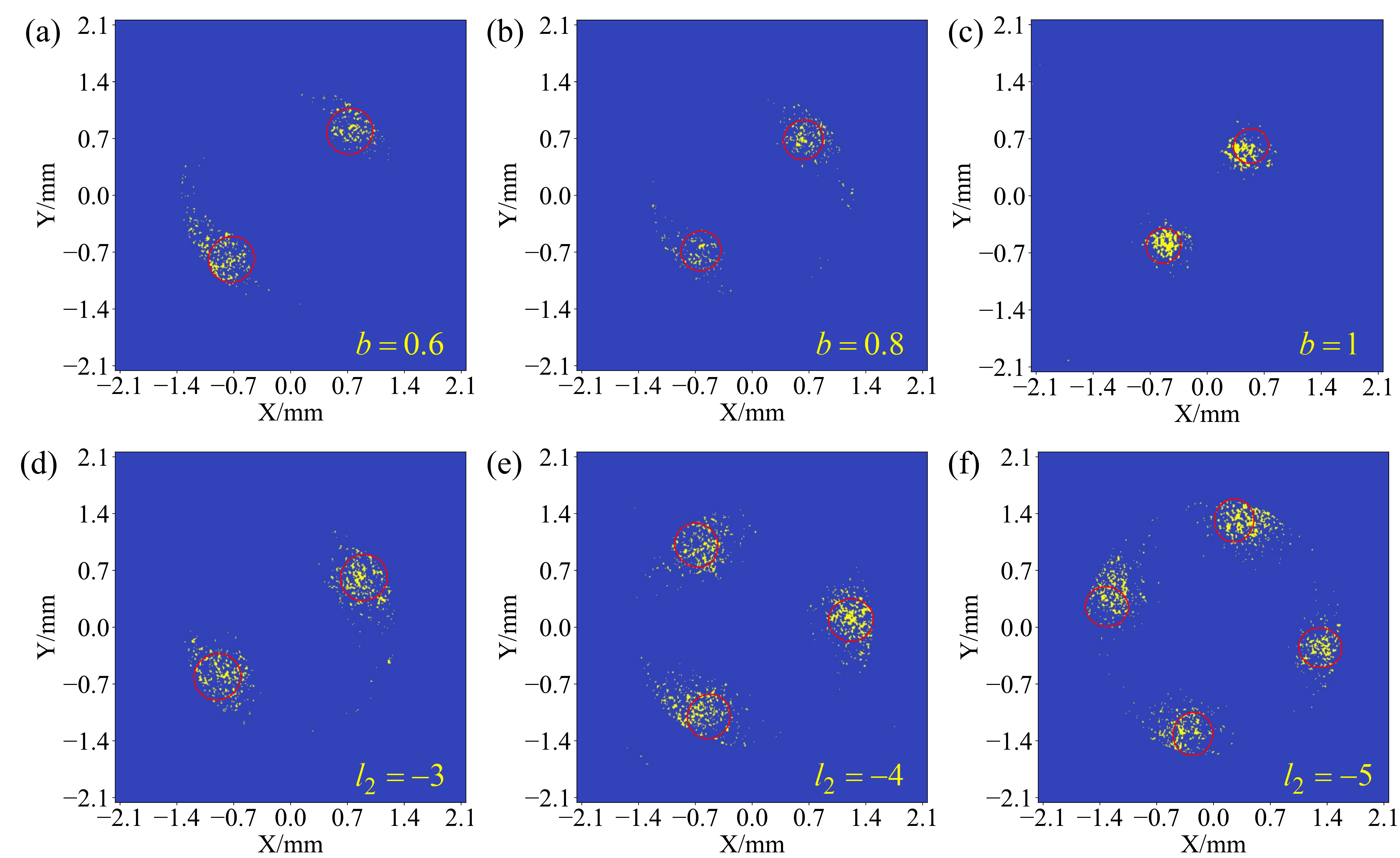}
	\caption{The effect of mode ratio $b$, propagation distance $z$ and OAM index $l_2$  on the distribution of AB. The distribution of AB when the photon is prepared in the superposition of two LG modes with $l_{1}=-1$, $l_{2}=-3$, and $z=1755$ mm is shown in (a) for $b=0.6$, (b) for $b=0.8$, and (c) for $b=1.0$. The distribution of AB when the photon is prepared in the superposition of two LG modes with $l_{1}=-1$, $b=1.0$, and $z=2805$ mm is shown in (d) for $l_{2}=-3$, (e) for $l_{2}=-4$, and (f) for $l_{2}=-5$. The yellow shaded regions represent the experimental AB regions, while the regions enclosed by red curves represent the theoretical AB regions. }
	\label{effect}
\end{figure}

We further investigate the effect of the mode ratio $b$,  the propagation distance $z$ and the OAM index $l_2$  on the distribution of AB. The results are shown in Fig. \ref{effect}. Clearly, as the mode ratio $b$ increases from 0.6 to 1.0, each AB region contracts and shifts inward along the radial direction. However, as the propagation distance $z$ increases from $1755$ mm to $2805$ mm, each AB region expands, moving outward and rotating clockwise. Additionally, we fix the OAM index $l_1$ at $-1$ while gradually decreasing the OAM index $l_2$ from $-3$ to $-5$. Obviously, the number of AB regions increases from 2 to 4, consistently corresponding to the absolute value of the difference between these two OAM indices.

\section{Conclusion}

In summary, with the help of weak measurement technology, we explored the phenomenon of AB within a single-photon system, both theoretically and experimentally. The results indicate that a photon prepared in the superposition state of two LG modes with negative OAM indices exhibits positive azimuthal wavevector components  
in certain low probability density regions. This suggests that AB occurs in regions of destructive interference where the interference visibility exceeds the threshold. The number of AB regions equals the absolute value of the difference between these two OAM indices. Additionally, as the mode ratio decreases and the propagation distance increases, each AB region expands and shifts outward along the radial direction.  By further analyzing the photon trajectory, we found that the photon can periodically enter and exit the AB region during propagation, and the closer it is to the phase singularity, the higher the frequency becomes.

Our method of measuring momentum eliminates the need to incorporate a slit or lenslet array in front of the photon detector. This not only simplifies the setup but also enhances the spatial resolution to the single pixel level. Additionally, by replacing the classical optical beam employed in previous studies with a single photon source, we take a significant step towards observing QB in an authentic quantum system.
The observed AB phenomenon, a characteristic of superoscillations, can not only be utilized to develop superoscillatory lenses, imaging, and metrology techniques \cite{zheludev2021b}, but also to evaluate quantum advantages in transportation tasks \cite{trillo2023}.
If the single-photon source is replaced with an entangled photon source, the mode ratio between the two superposition states can be nonlocally changed, then our method can be further employed to observe the nonlocal effect of QB and investigate the relation between QB and nonlocal correlations.

\begin{backmatter}
\bmsection{Funding}
This work was supported by the Natural Science Foundation of Shandong Province of China (Grant No. ZR2021ZD19), the Fundamental Research Funds for the Central Universities (Grants No. 202364008), and the Young Talents Project at Ocean University of China (Grant No. 861901013107).

\bmsection{Disclosures} The authors declare no conflicts of interest.

\bmsection{Data availability} The data sets generated during the current study are available from the corresponding author on reasonable request.

\bmsection{Supplemental document}
See Supplement 1 for supporting content.
\end{backmatter}

\bibliography{azimuthal_backflow}
\bibliographyfullrefs{azimuthal_backflow}


\ifthenelse{\equal{\journalref}{aop}}{%
\section*{Author Biographies}
\begingroup
\setlength\intextsep{0pt}
\begin{minipage}[t][6.3cm][t]{1.0\textwidth} 
  \begin{wrapfigure}{L}{0.25\textwidth}
    \includegraphics[width=0.25\textwidth]{john_smith.eps}
  \end{wrapfigure}
  \noindent
  {\bfseries John Smith} received his BSc (Mathematics) in 2000 from The University of Maryland. His research interests include lasers and optics.
\end{minipage}
\begin{minipage}{1.0\textwidth}
  \begin{wrapfigure}{L}{0.25\textwidth}
    \includegraphics[width=0.25\textwidth]{alice_smith.eps}
  \end{wrapfigure}
  \noindent
  {\bfseries Alice Smith} also received her BSc (Mathematics) in 2000 from The University of Maryland. Her research interests also include lasers and optics.
\end{minipage}
\endgroup
}{}

\end{document}